# Exploring Arsenic danger awareness in the Polish Copper Basin via network simulation – preliminary results

Andrzej Buda, Andrzej Jarynowski

Institute of Interdisciplinary Research, Głogów, Poland

andrzej.jarynowski@sociology.su.se, andrzejbudda@gmail.com

**Abstract**. Information spread plays crucial role in risk management in case of environmental danger. The structure of local society may be well described by social network analysis. We have investigated the role of hubs within that concept. In case of danger, there are two different strategies of information management: 1) information spread that leads to awareness of the whole society 2) keeping the whole information in secret under control that leads to a partial social awareness, available to a small number of people only. In our model, the probability of information spread between two nodes is inverse proportional to connectivity of the next node because people who have a lot of connections are more immune. We have applied agent-based modelling on Barabasi-Albert networks to explore various scenarios of information spread. We have considered recent arsenic environmental danger in Głogów and Legnica (Copper Basin) according to the official available data (2015-2016) from social network analysis point of view. We have considered various levels of environmental danger. Despite blocking of information by hubs, the successful information spread is possible when levels of danger are high enough. The perception and impression of information spread by society is also.

## 1    Arsenic

Arsenic substance is very dangerous for health including a carcinogenic effect in human body. It causes the development of skin tumors, lung, liver. It may affect skin lesions, such as actinic pigmentation, inflammation. There are restrictions in quantity of arsenic concentrations in the air (6 nanogram per cubic meter). The cities of Głogów and Legnica are situated in southwestern Poland (Lower Silesian Voivodeship) next to 2 copper smelters (owned by KGHM) and contain around 100-200 k citizens. According to the Provincial Environmental Protection Inspectorate, the arsenic level in Głogów and Legnica areas exceed few time above norm (measurement has been performed since February 2015).

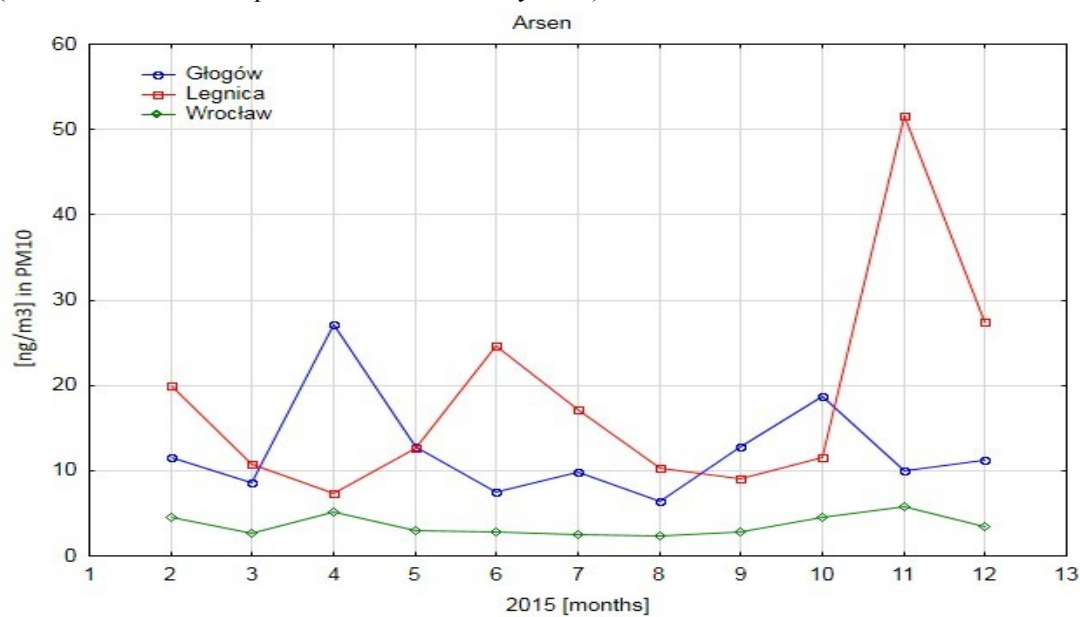

Fig 1.  Arsenic level in Głogów and Legnica in comparison with the capitol of Lower Silesia Region – Wrocław

## 2     Social movements and social network analysis of environmental awareness

Social world deals, in particular, with connected systems endowed with network structure. Such a network, is a subject to certain structure often not known to agencies building it. Since the appearance of the interdisciplinary perspective known as Social Network Analysis [1] the number of researchers in the social and behavioral sciences who are familiar with graph theory language, algorithms and theorems has been systematically growing.

In network terminology for environmental issues:
- Node = individual components of a network (agent) e.g. Local media, Policy Makers, Activist, Pollution Emitters
- Edge = indirect link between components (referred to social networking as a relationship between two agents).
- Path = route taken across components to connect two nodes (it's very important issue, where we looking on influential spreader).

Typical network properties are also listed:
- Clusters: Clustering coefficient counts number of triangles in networks.
- Community's detection.
- Average (shortest) path length – important in flow of information's on network.
- Degree distribution- distribution of connections of nodes.
- Randomness: From Grid/lattice network (ordered structure), via Small-world network (a mix of order and randomness), to Random networks (usually Barabasi-Albert: BA with power low degree distribution or Erdos-Renyi: ER with exponential degree distribution).

We identify most important agents:
- Hub - the most direct connections in the network, making it the most active node in the network
- Bridge - situated between many important constituencies (path)

Following a discussion of few formal approaches to interacting systems with focus on the social change, we present the sociological terminology and few theorems. In social science there are three possible ways of explaining the success and failure of social movements: through the development of potent master frames and how they may be rendered impotent by sociopolitical changes; the presence/absence of societal strain, sometimes caused by economic hardships and relative deprivation, that may result in waves of social protest; and through the expansion or contraction of political opportunities such as the level of trust in established political institutions [2]. Critical phenomena in language of social movements is also a big field of qualitative research. On the other hand, the knowledge about critical phenomena in economy is based on time series and correlations between assets. Thus, there are many other helpful tools including life-time of correlations or the Hurst exponent [3] that may be extended to description and predictions of environmental phenomena (for instance: the Nile river behaviour). In our research, we focus on agent-based modeling applied to social networks. That is an explanation of using the terminology borrowed from the epidemiological modeling [4] and innovation diffusion [5]. Before we consider the Glogow case, let us show 2 more famous examples from Easter Europe: Cracow and Chernobyl.

**Air Pollution and ban for solid fuels in Krakow**

Air pollution in Krakow poses a significant danger to human health and life (alarm). Krakow is ranked in top 10 most polluted cities in European Union in many dimensions. Through the years (2012-2016) a public opinion appears to have mobilised by "Krakowski Alarm Smogowy" – a social movement fighting with pollution emitters. The success of this movement is a combination of successful spread of viral information (mostly in social media) and governmental substitutes fueling the process.

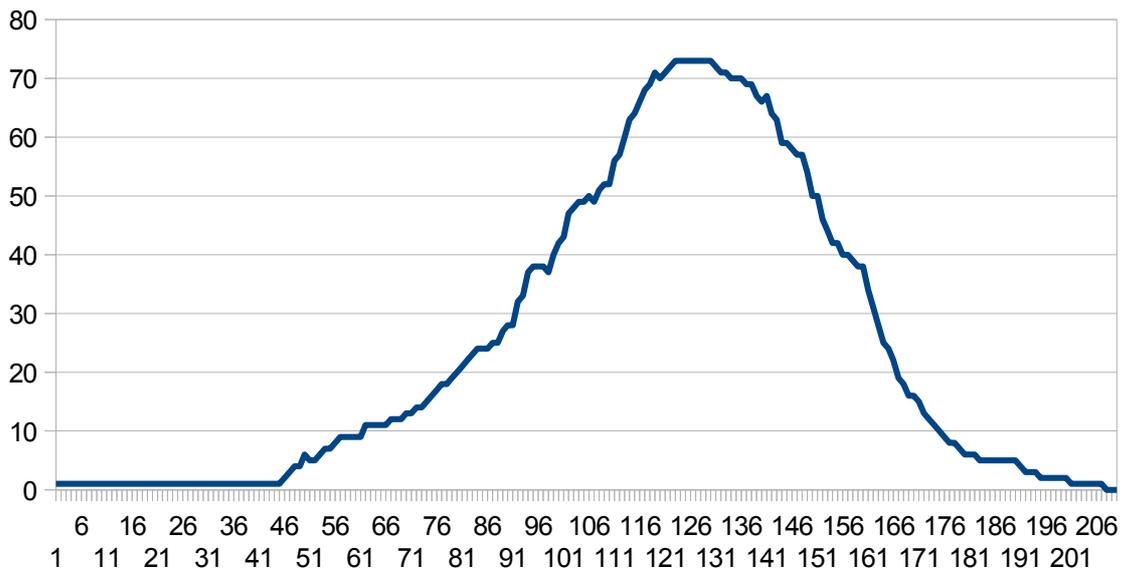

Fig 2. The average number of infected nodes as a result of simulation in Cracow like scenario. Aggressiveness a =0.1, m=50 (described in next section)

**Chernobyl disaster and the spiral of silence**

After the Chernobyl Nuclear Power Plant catastrophe, Soviet Authorities declined that huge radioactive could badly affect millions of people in Europe. On April the 28[th] (1986) The Soviet TV "Wremia" journalists made a comment "The effects of the accident are being remedied". The information spread was stopped at the beginning and no assistance had been provided for citizen of Communistic Countries until the Western authorities spread out knowledge using their own channels of propaganda.

People engaging in online networking tend to gravitate to groups agreeing with their own views and listen to authorities which can be network hub of bridge. Common knowledge (understanding common goals as environment) is a necessary condition for undertaking collective actions. Revolutions start when there is a large disparity between expectations and capabilities to satisfy them.

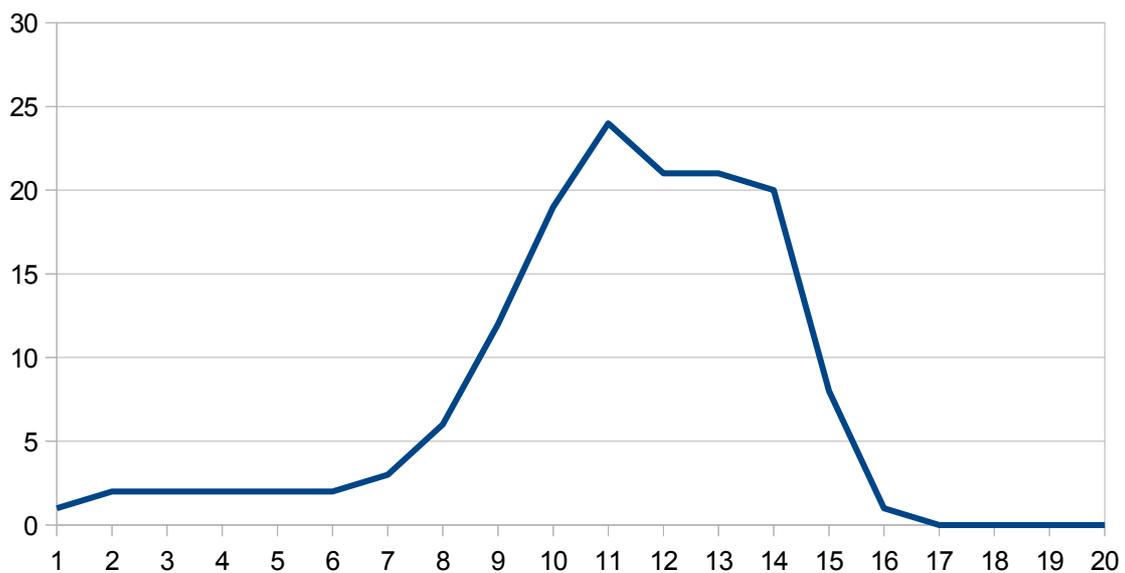

Fig 3. The average number of infected nodes as a result of simulation in Cherobyl like scenario. Aggressiveness a =1, m=2 (described in next section)

We claim, that network hubs are extremely important in broadcasting information and in essence of environmental issues, they are usually against spreading the knowledge because of incentives in keeping it secret.

## 3   Model

It is also relevant to give a sneak preview of an epidemiologic model of information spread, based on scal-free network. Each of local institution is represented by a number of nodes proportional to the size of a local actor [6]. The total number of nodes in Głogów network has been normalized to be equal 100. We could define viruses as an information that affect one actor after another [7]. Firstly, we choose a node as a source of infection randomly.

The probability p of affecting the nearest neighbours is proportional to:
1) the aggressiveness *a* of a product
2) *k* - number of infected neighbours after every consecutive step (*p* is also inversely proportional to the total number of nearest neighbours). This impose, that hubs are more immune than rest of society.

Before running this simulation, it is also necessary to define *m* - the time of disease for a node [in steps or months] as a parameter associated with local discourse and level of environmental danger.

If we consider the local scene of government, journalist, activist as well as pollution producers, we can settle *m*=2 because in each step agent can decide if he engage in action and share their opinion to their nearest neighbours next month. So, this innovation diffusion step takes 2 months only. It is clearly visible that aggressiveness of an information must be efficient enough to infect the whole society.

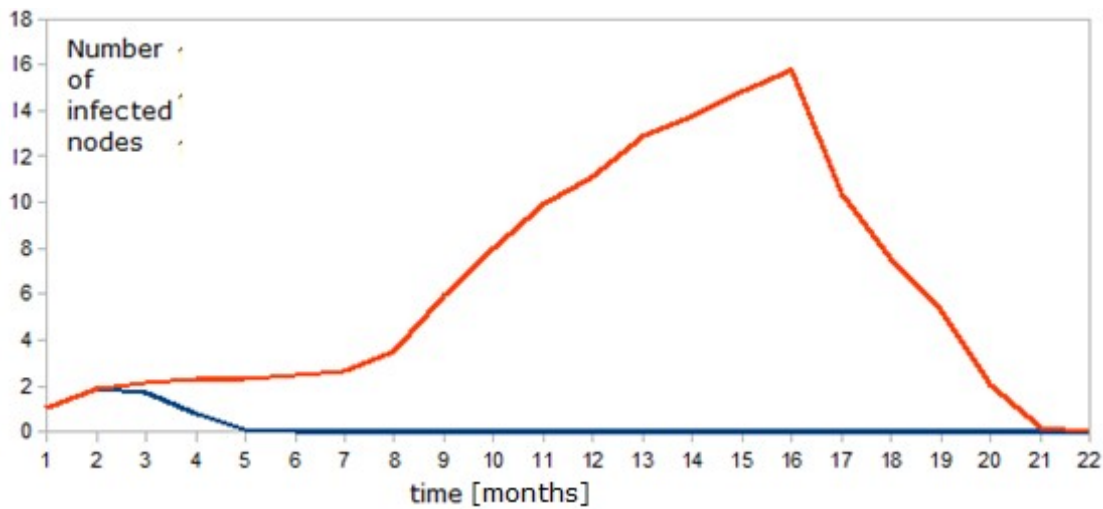

Fig. 4. The average number of infected nodes in Głogów network as a result of simulation. Aggressiveness a = 0.85 (upper line) is efficient enough to infect all the nodes in 21 months. However, if the virus has a = 0.8, the spread of popularity will decay completely in 5 months. We consider the economic circumstances and influences only (m = 2).

On the other hand, many processes need building social movements are it need much more time. Thus, it is reasonable to consider m bigger than 2. The epidemic will spread easy to all other nodes. Our aim in future works is to consider more complicated networks and modern marketing strategies (with or without medial support) applied to local scene. Of course, keeping the nodes of local authorities in state of disease for m bigger than 2 extends story life-cycles, but it involves more costs and the return in various diffusion scenarios may not be optimal for social movements. On the other hand, medial support (*m* bigger than 2) is more than just the economy. It gives a benefit because all stages of product life-cycles are extended. This epidemiologic model of an environmental awareness propagation is worth to investigate because of complete data sets with high accuracy and relevant structural marketing purposes.

## 4   Conclusions and further work

According to our agent-based model, we predict that people involved in social network in Glogow will be less and less interested in arsenic danger in 2017. This issue, however, is the main topic in local news because people are more and more aware. On the other hand, the solution of this problem is necessary for all the community, including The KGHM Company [8]. Thus, according to our predictions, the average number of infected nodes of this social network is going to decrease fast after solving the problem. Our agent-based epidemiologic model of environmental awareness propagation is worth to investigate because of complete data sets with high accuracy and relevant structural marketing purposes. In our future work, we are going to extend our model to the health parameters of all citizens in Glogow according to the official medical base (Narodowy Fundusz Zdrowia). The source of the pollution may be detected by spatial distribution of citizens diseases.

## Literature

[1] **Jarynowski, A.** *Obliczeniowe nauki społeczne w praktyce.* Wrocław : WN, 2014.

[2] **Rydgren, J.** Is extreme right-wing populism contagious? Explaining the emergence of a new party family. *European Journal of Political Research*,*44*(3), 413-437. 2005.

[3] **Buda, A. and Jarynowski, A.** *Lifetime of Correlation and its applications.* Wroclaw : WN, 2011.

[4] **Rogers, E.M.** *Diffusion of Innovations.* New York : Free Press, 2005.

[5] **Anderson RM and May RM,** Infectious Diseases of Humans: Dynamics and Control, Oxford: Oxford University Press, 1992.

[6] **Castells, M.** The rise of the network society: The information age: Economy, society, and culture. Vol. 1. John Wiley & Sons, 2011.

[7] **Leskovec, J. Adamic, L and Huberman, B.** "The dynamics of viral marketing," ACM Trans. Web, vol. May, 2007.

[8] **Zbieg, A.,** et al. "Identyfikacja wartości w przedsiębiorstwie górniczym w strategicznej koncepcji zarządzania przez wartości." *Wiadomości Górnicze* 67. 2016.